\documentclass[a4paper,fleqn,usenatbib]{mnras}

\usepackage{newtxtext,newtxmath}

\usepackage[T1]{fontenc}
\usepackage{ae,aecompl}

\usepackage{graphicx}	
\usepackage{amsmath}	
\usepackage{amssymb}	

\newcommand{\kms}{ km s$^{-1}$}

\newcommand{\II}{~{\sc ii}}
\newcommand{\I}{~{\sc i}}

\title{Quasar 2175 \AA$ $ dust absorbers II: Correlation analysis and relationship with other absorption line systems}
\author[Jingzhe Ma, Jian Ge, J. Xavier Prochaska, Shaohua Zhang, Tuo Ji, Yinan Zhao, Hongyan Zhou, Honglin Lu, Donald P. Schneider]{Jingzhe Ma$^{1}$\thanks{E-mail:jingzhema@ufl.edu (JM)}, Jian Ge$^{1}$, J. Xavier Prochaska$^{2}$, Shaohua Zhang$^{3}$, Tuo Ji$^{3}$, 
\newauthor Yinan Zhao$^{1}$, Hongyan Zhou$^{3}$, Honglin Lu$^{4}$, Donald P. Schneider$^{5,}$$^{6}$\\
$^{1}$Department of Astronomy, University of Florida, 211 Bryant Space Science Center, Gainesville, 32611, USA\\
$^{2}$Department of Astronomy and Astrophysics, UCO/Lick Observatory, University of California, 1156 High Street, Santa Cruz, 95064, USA\\
$^{3}$Polar Research Institute of China, 451 Jinqiao Road, Pudong, Shanghai 200136, China\\
$^{4}$Department of Astronomy, University of Sciences and Technology of China, Chinese Academy of Sciences, Hefei, Anhui 230026 China\\
$^{5}$Department of Astronomy and Astrophysics, The Pennsylvania State University, University Park, PA 16802, USA\\
$^{6}$Institute for Gravitation and the Cosmos, The Pennsylvania State University, University Park, PA 16802, USA\\
}

\begin{document}
\maketitle

\begin{abstract}

We present the cold neutral content (H\I{} and C\I{} gas) of 13 quasar 2175 \AA$ $ dust absorbers (2DAs) at $z$ = 1.6 - 2.5 to investigate the correlation between the presence of the UV extinction bump with other physical characteristics. These 2DAs were initially selected from the Sloan Digital Sky Surveys I - III and followed up with the Keck-II telescope and the Multiple Mirror Telescope as detailed in our Paper I. We perform a correlation analysis between metallicity, redshift, depletion level, velocity width, and explore relationships between 2DAs and other absorption line systems. The 2DAs on average have higher metallicity, higher depletion levels, and larger velocity widths than Damped Lyman-$\alpha$ absorbers (DLAs) or subDLAs. The correlation between [Zn/H] and [Fe/Zn] or [Zn/H] and log$\Delta$V$_{90}$ can be used as alternative stellar mass estimators based on the well-established mass-metallicity relation. The estimated stellar masses of the 2DAs in this sample are in the range of $\sim$ 10$^9$ to $\sim$2 $\times$ 10$^{11}$ $M_{\odot}$ with a median value of $\sim$2 $\times$ 10$^{10}$ $M_{\odot}$. The relationship with other quasar absorption line systems can be described as (1) 2DAs are a subset of Mg\II{} and Fe\II{} absorbers, (2) 2DAs are preferentially metal-strong DLAs/subDLAs, (3) More importantly, all of the 2DAs show C\I{} detections with logN(C\I) $>$ 14.0 cm$^{-2}$, (4) 2DAs can be used as molecular gas tracers. Their host galaxies are likely to be chemically enriched, evolved, massive (more massive than typical DLA/subDLA galaxies), and presumably star-forming galaxies.

\end{abstract}


\section{Introduction}

\begin{table*}
\centering
\caption{Properties of the quasar 2175 \AA$ $ absorbers. [Fe/Zn], $A_{\rm V}$, and $A_{\rm bump}$ are measured in the same manner as described in \citealt{Ma2017}. }
\begin{tabular}{cccccccccc}
\hline\hline
Sourcename        & Telescope & $z_{em}$  & $z_{abs}$  & log$N{\rm (HI)}$  & [Zn/H]  & [Fe/Zn]  & W$_{r}$($\lambda$1560)&  $A_{\rm V}$ & $A_{\rm bump}$\\  
\hline
J0745+4554  &  Keck  & 2.1998   & 1.8612   &                               &                              & -1.41 $\pm$ 0.09  & 0.941 $\pm$ 0.074  &0.95 $^{+ 0.09}_{- 0.18}$  & 0.676 $^{+ 0.296}_{- 0.086}$\\[0.07cm]
J1006+1538  & Keck  &  2.1817   & 2.2062   & 20.00 $\pm$ 0.15  & 0.65 $\pm$ 0.26  & -1.19 $\pm$ 0.24  & 0.402 $\pm$ 0.045 &0.41 $^{+ 0.07}_{- 0.17}$ & 0.433 $^{+ 0.362}_{- 0.086}$\\[0.07cm]
J1047+3423  & Keck  &  1.6800    & 1.6685   & 20.05 $\pm$ 0.20 & 0.54 $\pm$ 0.22  & -1.40 $\pm$ 0.09 & 0.415 $\pm$ 0.103 & 0.61 $^{+ 0.08}_{- 0.18}$  & 0.194 $^{+ 0.097}_{- 0.027}$ \\[0.07cm]
J1130+1850   & Keck  &  2.7536   & 2.0119   & 21.10 $\pm$ 0.30  & 0.04 $\pm$ 0.30  & -0.73 $\pm$ 0.03 & 0.602 $\pm$ 0.160 &0.49 $^{+ 0.04}_{- 0.18}$ & 0.641 $^{+ 0.431}_{- 0.115}$\\[0.07cm]
J1141+4442   & MMT  &  1.9637   & 1.9016   & 20.85 $\pm$ 0.15 & -0.09 $\pm$ 0.19  & -1.04 $\pm$ 0.12 & 0.404 $\pm$ 0.096 & 0.24 $^{+ 0.11}_{- 0.16}$  & 0.574 $^{+ 0.373}_{- 0.164}$  \\[0.07cm]
J1157+6155   & MMT  &   2.5120  &  2.4596  & 21.80 $\pm$ 0.20 & -0.27 $\pm$ 0.23  &-1.40 $\pm$ 0.22 & 1.220 $\pm$ 0.140 &1.04 $^{+ 0.11}_{- 0.16}$& 0.377 $^{+ 0.376}_{- 0.112}$   \\[0.07cm]
J1209+6717   & MMT  &  2.0300   & 1.8425  &  20.25 $\pm$ 0.20 & $<$ 0.41    & $>$ -1.33                       & 0.391 $\pm$ 0.140                      & 0.18 $^{+ 0.07}_{- 0.10}$& 0.451 $^{+ 0.261}_{- 0.152}$\\[0.07cm]
J1211+0833   & Keck  &   2.4828  & 2.1166   & 21.00 $\pm$ 0.20 & 0.02 $\pm$ 0.21   & -1.75 $\pm$ 0.07 & 0.898 $\pm$ 0.145 & 0.80 $^{+ 0.05}_{- 0.18}$     & 1.107 $^{+ 0.636}_{- 0.169}$ \\[0.07cm]
J1321+2135   &Keck  &   2.4113   & 2.1253   & 21.55 $\pm$ 0.20 & -0.50 $\pm$ 0.20  & -0.71 $\pm$ 0.04& 0.263 $\pm$ 0.119 &0.39 $^{+ 0.03}_{- 0.18}$     & 0.328 $^{+ 0.218}_{- 0.059}$ \\[0.07cm]
J1524+1030   & MMT  &  2.0620   & 1.9395   & 21.45 $\pm$ 0.10 &-0.45 $\pm$ 0.10  & -1.24 $\pm$ 0.07 & 0.194 $\pm$ 0.040 & 0.50 $^{+ 0.07}_{- 0.18}$& 0.132 $^{+ 0.095}_{- 0.071}$ \\[0.07cm]
J1531+2403   &Keck  &   2.5256   & 2.0022   & 20.20 $\pm$ 0.25 & 0.45 $\pm$ 0.21  & -1.25 $\pm$ 0.06 & 0.374 $\pm$ 0.089&0.44 $^{+ 0.03}_{- 0.19}$ & 0.562 $^{+ 0.295}_{- 0.079}$ \\[0.07cm]
J1705+3543   & MMT  &  2.0100   &  2.0377  & 20.62 $\pm$ 0.12 & 0.14 $\pm$ 0.14  & -1.63 $\pm$ 0.08  & 0.700 $\pm$ 0.080&0.25 $^{+ 0.04}_{- 0.18}$& 0.416 $^{+ 0.336}_{- 0.114}$ \\[0.07cm]
J1737+4406   &Keck  &  1.9564    &  1.6135  &                              &                             & -1.00 $\pm$ 0.08 & 0.271 $\pm$ 0.168&0.50 $^{+ 0.07}_{- 0.19}$ & 0.686 $^{+ 0.338}_{- 0.088}$ \\[0.07cm]
\hline
\end{tabular}
\label{tab:table1}
\end{table*}

Quasar 2175 \AA$ $ dust absorbers (2DAs) are a population of quasar absorption line systems identified by the broad absorption feature centered around rest-frame 2175 \AA$ $ (e.g., \citealt{Wucknitz2003,Wang2004,Junkkarinen2004,Srianand2008,Zhou2010,Jiang2010a,Jiang2010b,Jiang2011,Wang2012}), which is ubiquitously seen in the Milky Way (MW) extinction curves. These absorbers are excellent tracers of gas and dust properties, metal abundances, chemical evolution, physical conditions, as well as kinematics in the absorbing galaxies. 

The majority of the well-studied Damped Lyman-$\alpha$ systems (DLAs with logN(H\I) $\geq$ 20.3) have lower metallicity and lower dust content than modern galaxies and presumably the more massive galaxies at $z$ $>$ 1 (e.g., \citealt{Prochaska2007,Rafelski2012}). 2DAs offer a unique opportunity to study the interstellar medium (ISM) of putatively evolved systems (at least chemically). By studying the 2DAs, we are exploring the properties of the ISM in a distinct, likely more massive and evolved galaxy population. 

The first study of a sample of 2DAs with moderate to high resolution spectra (\citealt{Ma2017}; Paper I) reveals the presence of many metal absorption lines, mostly low-ionization lines (e.g. Zn\II, Fe\II, Mg\II, Si\II, Al\II, Mn\II, Cr\II, Ni\II, Ca\II, Ti\II) in the systems. In Paper I, we derived the relative metal abundances and thus depletion patterns based on the metal lines. The velocity profiles also provide insights into the kinematics of the absorbing gas. The 2DAs on average have higher depletion levels and larger velocity widths than DLAs or subDLAs (19.0 $<$ logN(H\I) $<$ 20.3; \citealt{Peroux2001}) in the literature. The high depletion levels confirm the presence of dust grains and therefore the 2175 \AA$ $ extinction bump. The larger velocity widths suggest that the 2DAs are expected to be more massive than typical DLA/subDLA galaxies, using velocity width as a proxy for mass. The 2DAs are more likely to be drawn from the same parent population as the metal-strong DLAs/subDLAs than normal DLAs/subDLAs \citep{Ma2017}. 

Along with the 2175 \AA$ $ bump, 2DAs are found to simultaneously harbor cold neutral (C\I, H\I, Cl\I) and/or molecular (CO and H$_2$) gas \citep{Eliasdottir2009, Noterdaeme2009, Prochaska2009, Ledoux2015, Ma2015}, which may serve as reservoirs for star formation in the host galaxies. A systematic search for C\I{} absorbers by \cite{Ledoux2015} shows that the 2175 \AA$ $ bump is present in $\sim$ 30\% of the C\I{} absorbers.

All the pieces of information suggest a correlation between the presence of the 2175 \AA$ $ bump and other properties. In the literature for DLAs, two correlations have been studied extensively. One is the evolution of metallicity as a function of redshift; metallicity gradually increases with decreasing redshift \citep{Prochaska2003,Kulkarni2005,Rafelski2012,Rafelski2014}. The second one is the correlation between the metallicity and kinematics \citep{Wolfe1998,Ledoux2006,Prochaska2008,Neeleman2013}, which is largely due to the underlying mass-metallicity relation that has been well-established in galaxies at high and low redshifts \citep{Tremonti2004,Savaglio2005,Erb2006,Maiolino2008,Moller2013,Neeleman2013}. The correlations between metallicity and other fundamental parameters such as stellar mass and star formation rate are key to understanding various galaxy populations \citep{Calura2009, Mannucci2010}.

In this paper, we perform a correlation analysis between metallicity, velocity width, redshift, depletion level, and additional quantities to reveal the underlying preferable conditions for the bump carriers to exist and the relationship of 2DAs with other absorption line systems. The outline of this paper is as follows. Section \ref{section2} introduces the sample of 2DAs in this work and associated observations. The measurements of H\I{} column densities and C\I{} equivalent widths are described in Sections \ref{section3} and \ref{section4}. Correlation analysis is presented in Section \ref{section5}. We discuss the relationships between 2DAs and other absorption line systems in Section \ref{section6}. The results are summarized in Section \ref{section7}.

\section{Sample Selection and Observations}
\label{section2}

\subsection{Sample Selection}

The 2DAs in this paper (Table \ref{tab:table1}) are drawn from the Sloan Digital Sky Surveys (SDSS) I-III \citep{York2000,Eisenstein2011} that have spectroscopic observations with Keck \citep{Ma2017} and Multiple Mirror Telescope (MMT; Hu et al. in prep). These are the sources whose redshifts are high enough such that neutral carbon lines (i.e. $\lambda$1560, 1656) and the Lyman-$\alpha$ absorption line are covered in SDSS, MMT, or Keck spectra.

How the best-fit extinction curves and bump parameters and uncertainties are determined is described in detail in Paper I. In a nutshell, we fit the model spectrum to the observed spectrum: a composite quasar spectrum is used as the intrinsic quasar spectrum and is reddened by a parameterized extinction curve to form the model spectrum. The bump strength measures the area under the extinction curve with a bump with respect to the one reddened by an underlying linear extinction only. This subset was selected for follow-up observations based on their bump strengths ($\gtrsim$ 0.2 $\micron$$^{-1}$) and accessibility with Keck/MMT. The total extinction, $A_V$, is estimated by assuming that extinction approaches zero as $\lambda$ $\to$ $+\infty$.

\subsection{Observations}

Eight absorbers in this work were followed up with the Echelle Imager and Spectrograph (ESI; \citealt{Sheinis2002}) on the Keck-II telescope on March 8 \& 9, 2013, using the 0.75$\arcsec$ slit. The wavelength coverage is 3900 - 11715 \AA{}  and the spectral resolution is $R =\lambda/\Delta\lambda$ $\sim$ 6000. We refer the reader to \cite{Ma2015} and \cite{Ma2017} for the observational details of individual sources. 

The MMT spectra were obtained using the Blue \& Red Channel Spectrographs \citep{Angel1979,Schmidt1989} with the 800 g mm$^{-1}$ and 500 g mm$^{-1}$ gratings on March 30, 2008. We use the 1$\arcsec$ slit to match the seeing conditions. The blue and red spectra cover a wavelength range of 3200 - 5200 \AA$ $ and 4400 - 7600 \AA$ $ at spectral resolutions of 1800 and 1600, respectively. The detailed description of the observations can be found in \cite{Zhou2010}, \cite{Wang2012}, \cite{Pan2017}, and Hu et al. in prep.

\section{H\I{} column densities}
\label{section3}

\begin{figure*}
{\includegraphics[width=7.5cm, height=4.5cm]{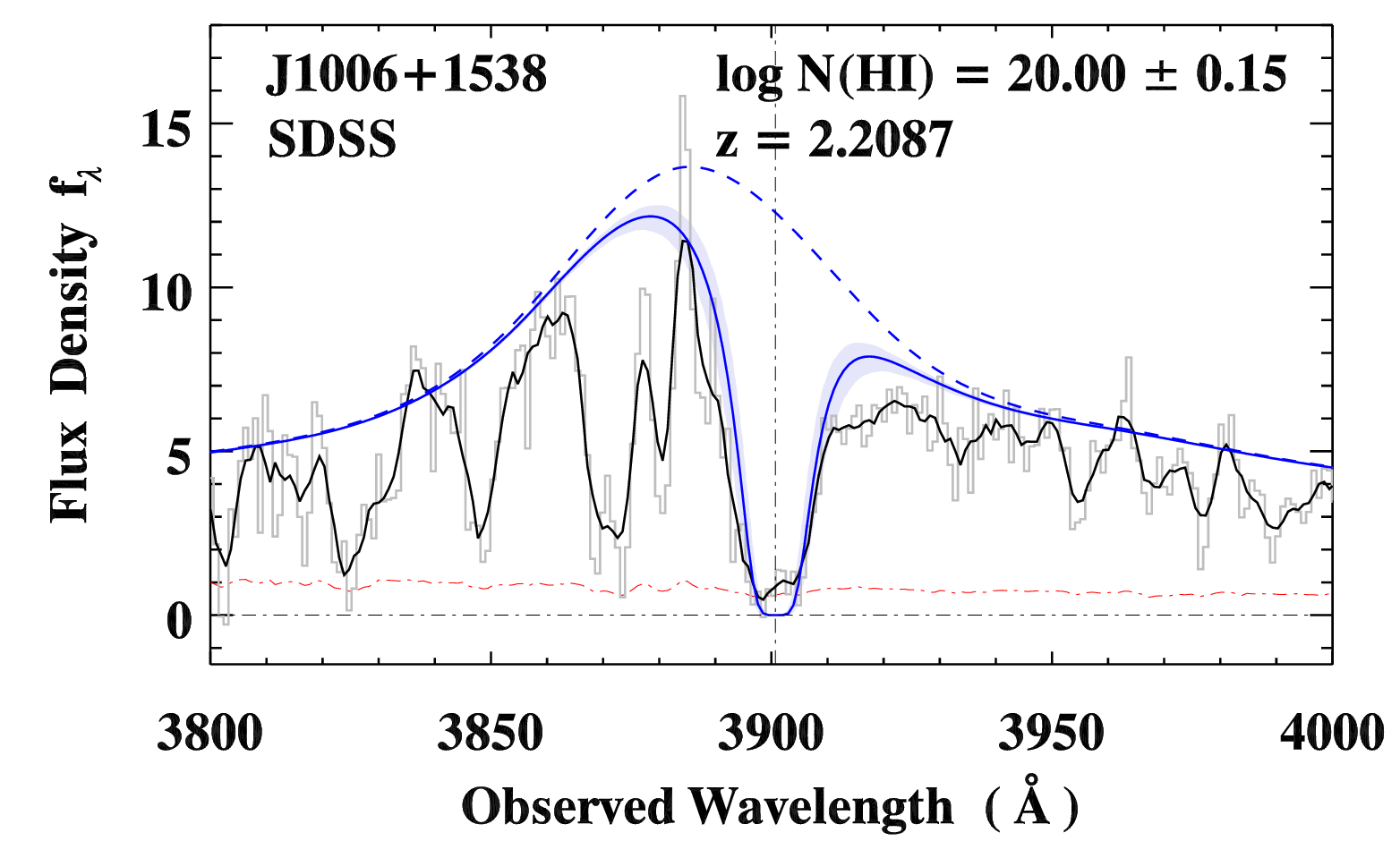}} 
{\includegraphics[width=7.5cm, height=4.5cm]{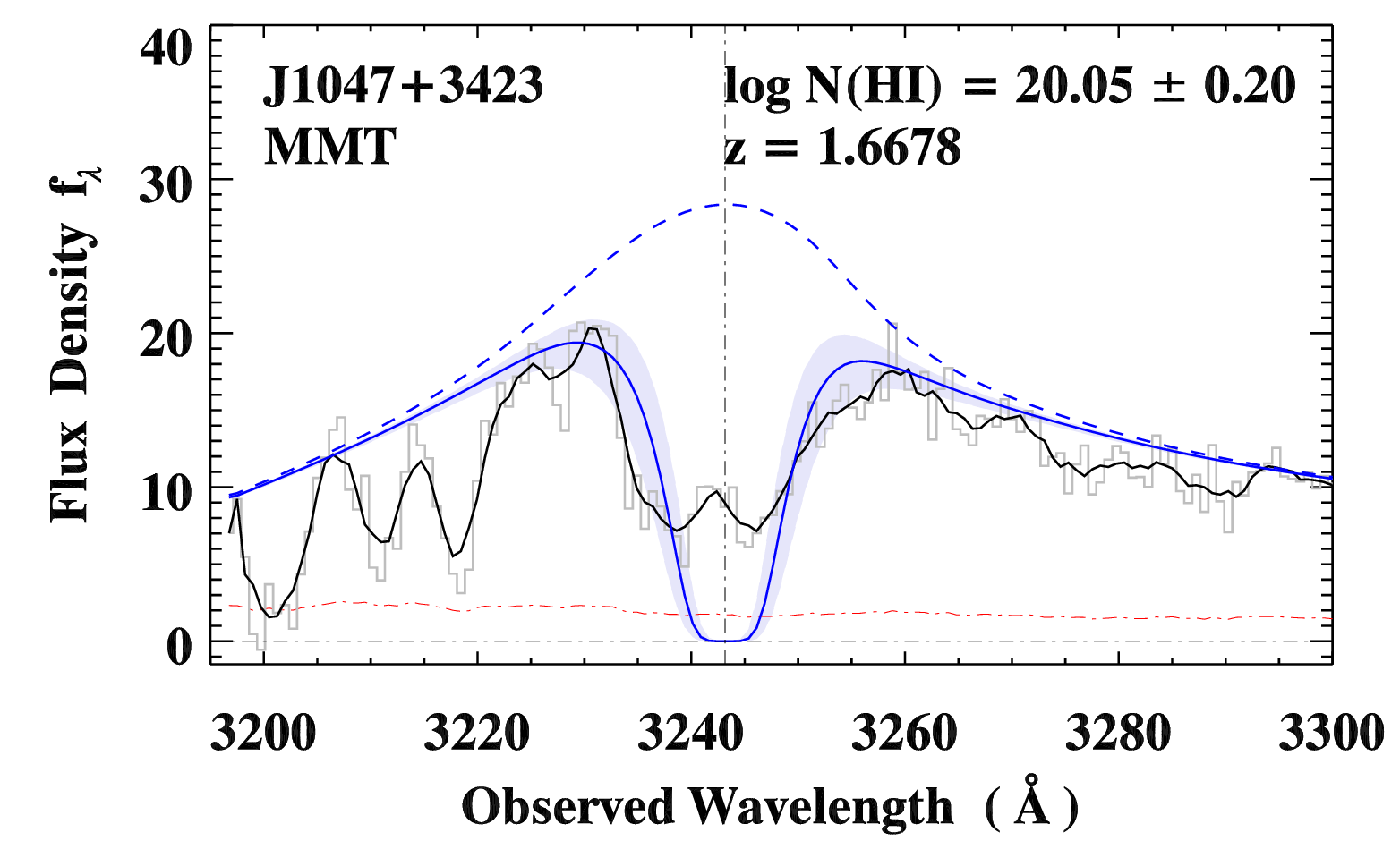}} 
{\includegraphics[width=7.5cm, height=4.5cm]{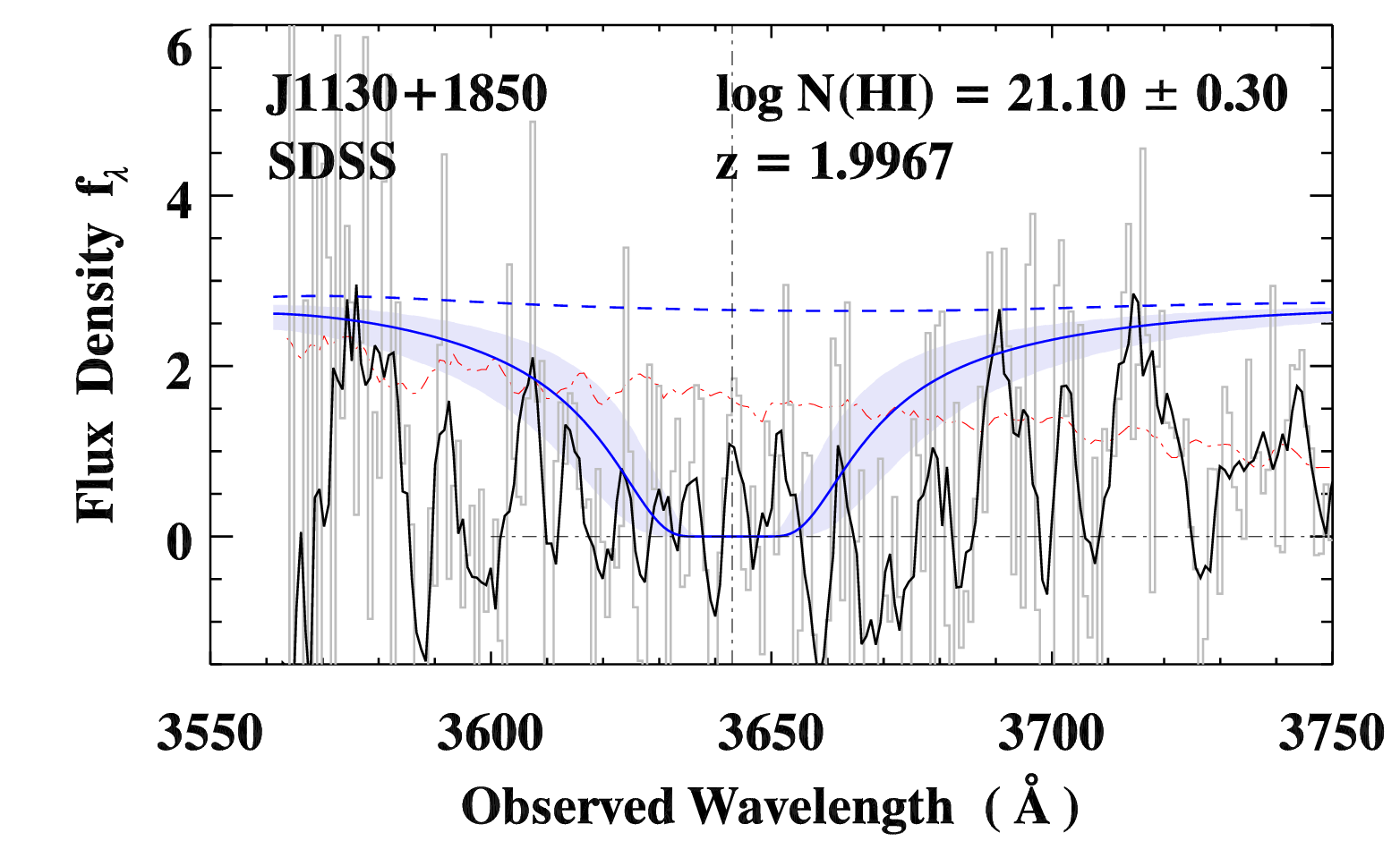}}   
{\includegraphics[width=7.5cm, height=4.5cm]{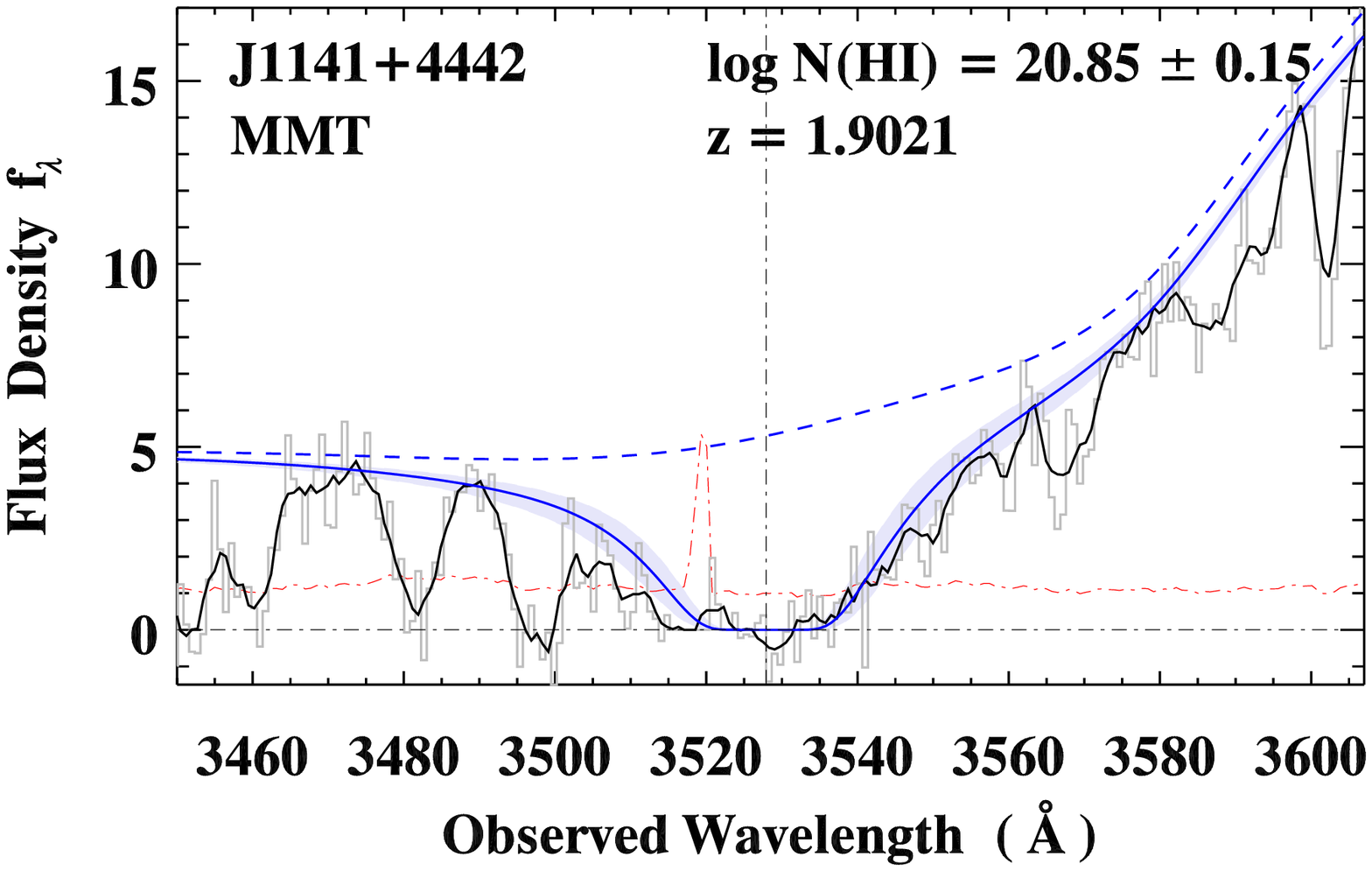}} 
{\includegraphics[width=7.5cm, height=4.5cm]{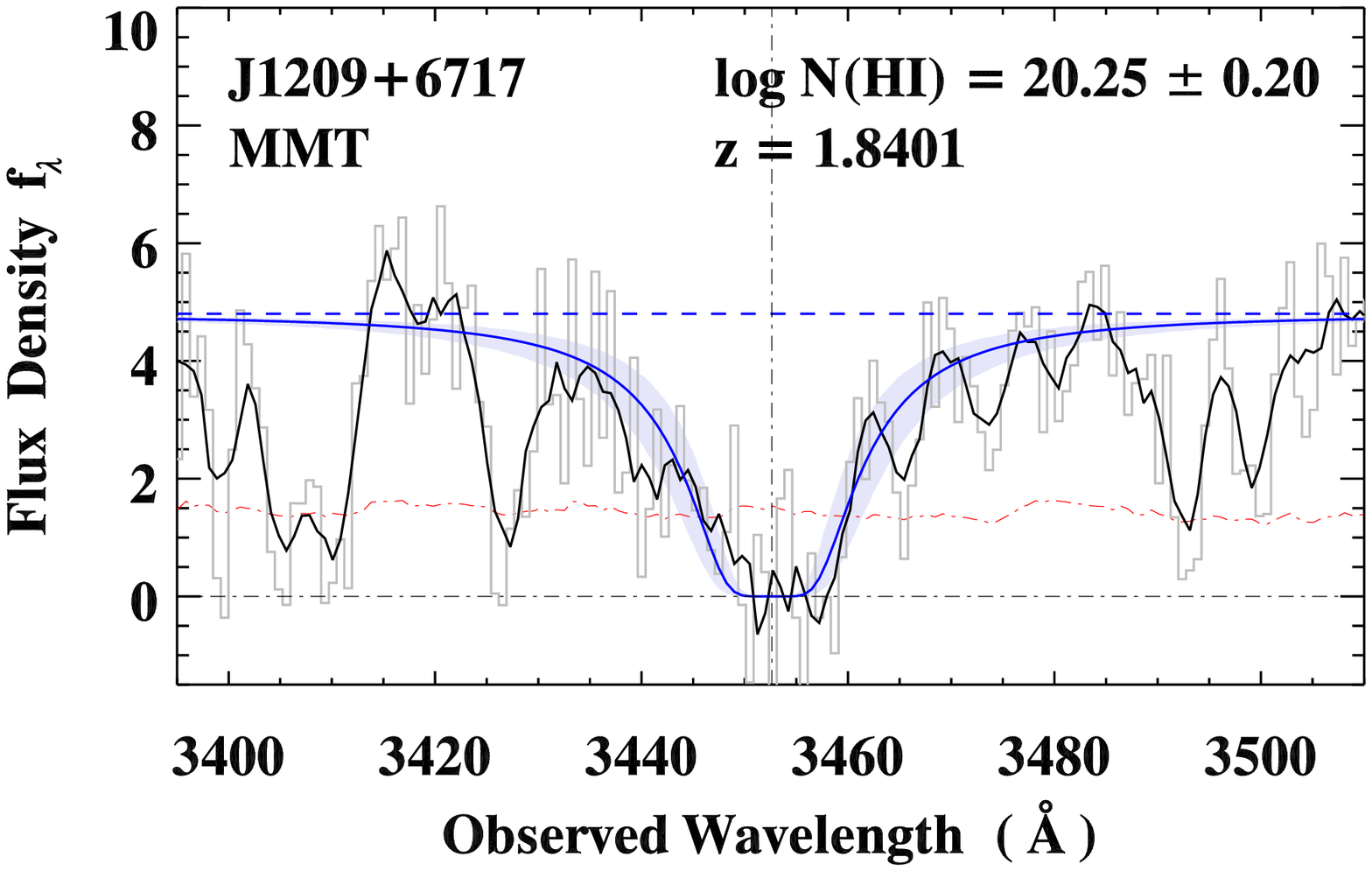}} 
{\includegraphics[width=7.5cm, height=4.5cm]{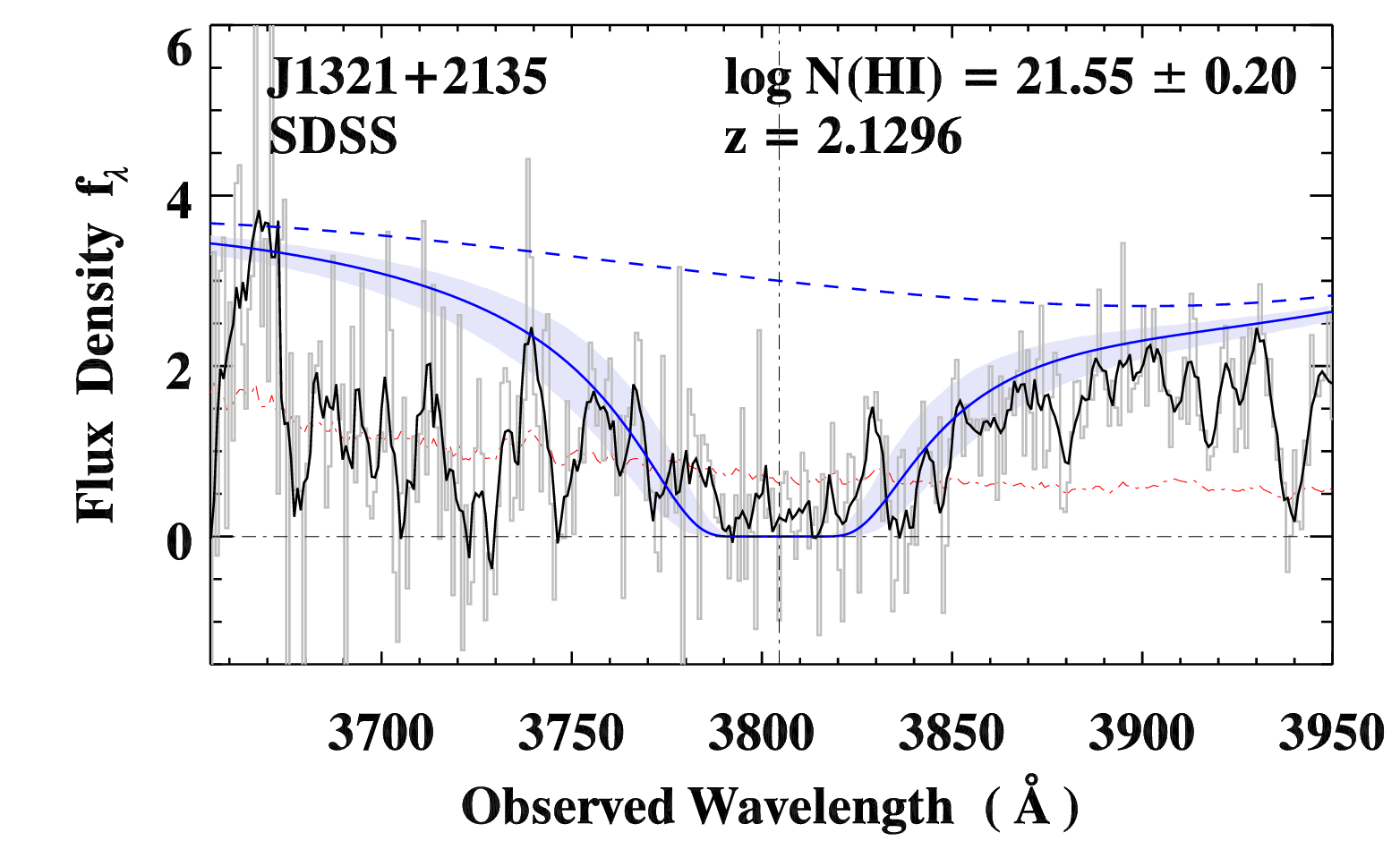}} 
{\includegraphics[width=7.5cm, height=4.5cm]{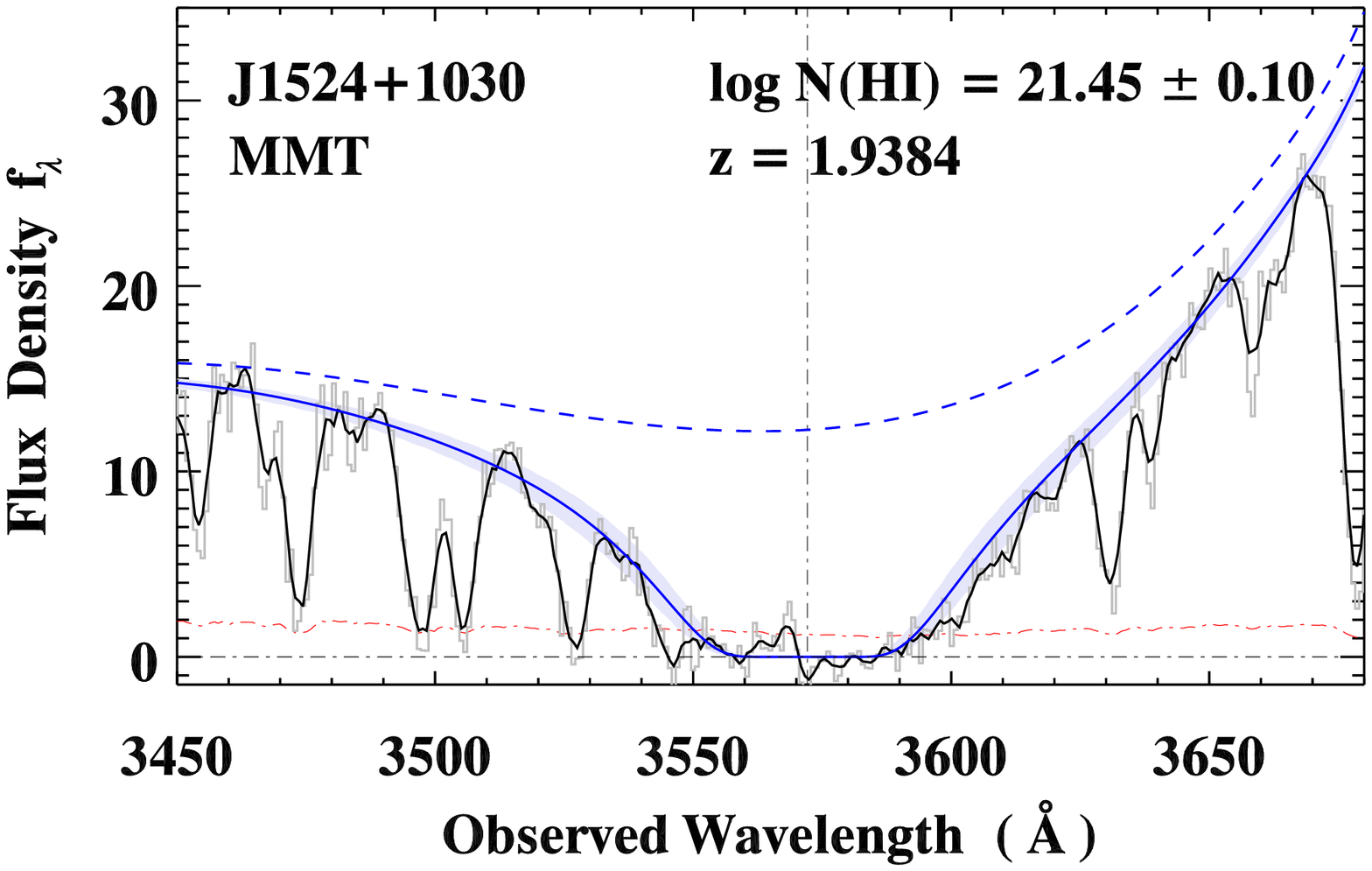}}
{\includegraphics[width=7.53cm, height=4.5cm]{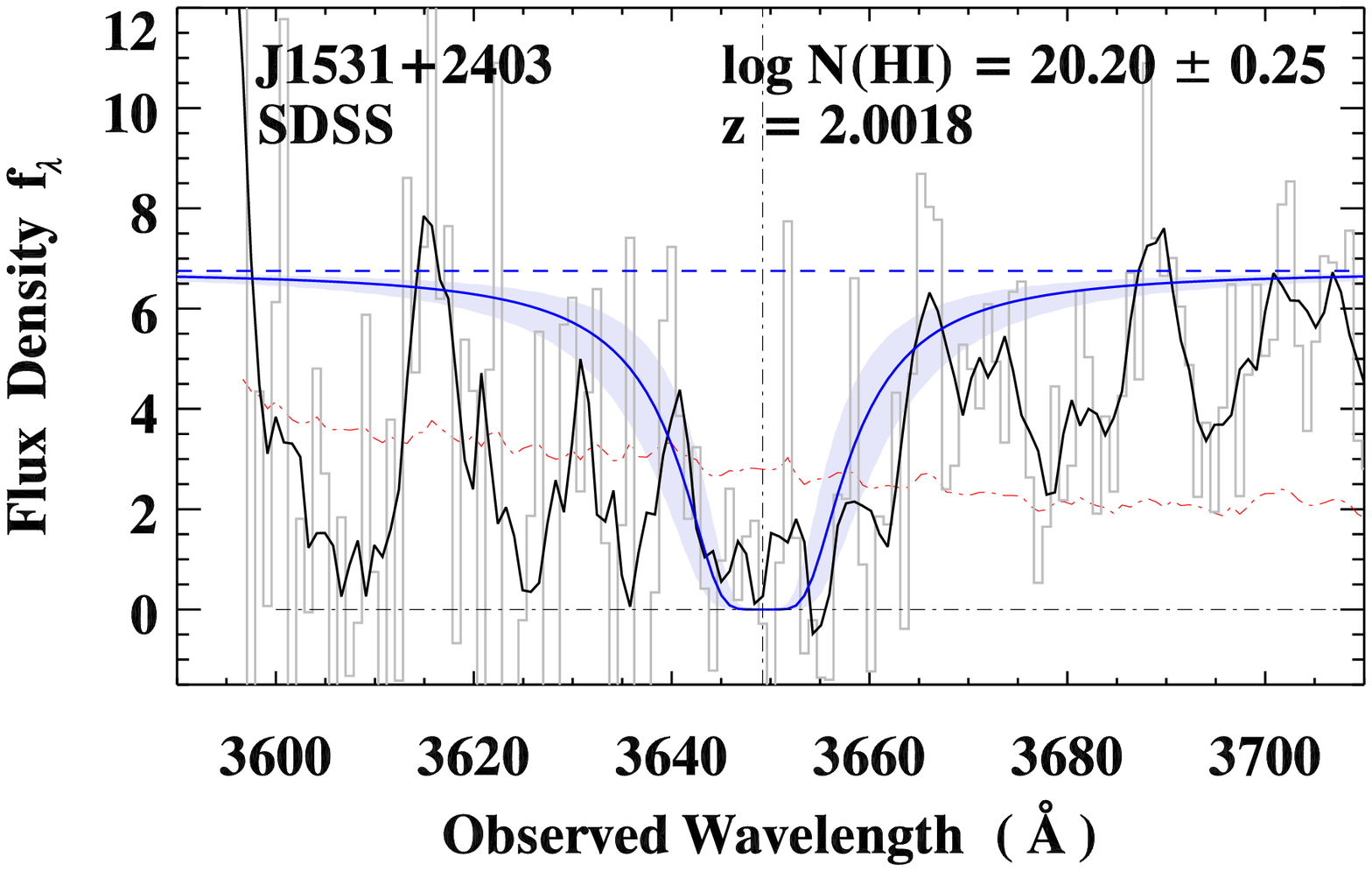}} 
\caption{DLA profile fitting on the SDSS or MMT spectra. $f_{\lambda}$ is in the units of 10$^{-17}$ erg s$^{-1}$ cm$^{-2}$ \AA$^{-1}$. The spectrum plotted in black is the original spectrum smoothed using 3-pixel boxcar and the red dotted line represents the associated flux errors. The blue line is the best-fit Voigt profile; the estimated 1 $\sigma$ uncertainty is denoted by the blue shade. The blue dashed line is the associated continuum. }
\label{fig:dla}
\end{figure*}

The Keck/ESI spectra do not cover the Lyman-$\alpha$ absorption line, therefore we use the SDSS spectra for the Keck absorbers. For the sources followed up with MMT, the blue channel spectra cover the Lyman-$\alpha$ absorption. Voigt profile fits are performed with x\_fitdla in the XIDL package\footnote{http://www.ucolick.org/\~xavier/IDL/index.html}. The best-fit Voigt profiles are shown in Figure \ref{fig:dla} with the 1 $\sigma$ uncertainty denoted by the shade. The error takes into account the systematic uncertainty due to placement of the continuum and small statistical uncertainty in the fitting. The absorber towards J1047+3423, whose absorption redshift is close to the quasar's emission redshift, exhibits residual Lyman-$\alpha$ emission superimposed on the absorption trough, a signature indicating that this 2DA could be intrinsic to the quasar (more discussion in Section \ref{sec:proximate}). 

We are able to derive the absolute metal abundances with the H\I{} column densities. The metal abundances are measured relative to solar values (photosphere abundances from \citealt{Asplund2009}) as [X/H] $\equiv$ log$N(\rm X)/N(\rm H)$ - log(X/H)$_{\sun}$ with the assumption that N(H) = N(H\I). The error on [X/H] is calculated from the errors on N(H\I) and N(X) through error propagation. In the following analysis, we assume the ionization corrections of metals are negligible for our absorbers. The assumption works well in the DLA regime due to the shielding effect of large neutral column density while in the subDLA regime the relatively low N(H\I) does not provide complete shielding and the ionized fraction of hydrogen varies \citep{Meiring2009, Lehner2014}. Although not negligible, the required ionization corrections to elemental abundances are often low ($<$ 0.3 dex; \citealt{Dessauges-Zavadsky2003}). We keep this uncertainty in mind in comparison with the measurements in the literature. 

\section{C\I{} equivalent widths}
\label{section4}

\begin{figure*}
{\includegraphics[width=12cm, height=15cm]{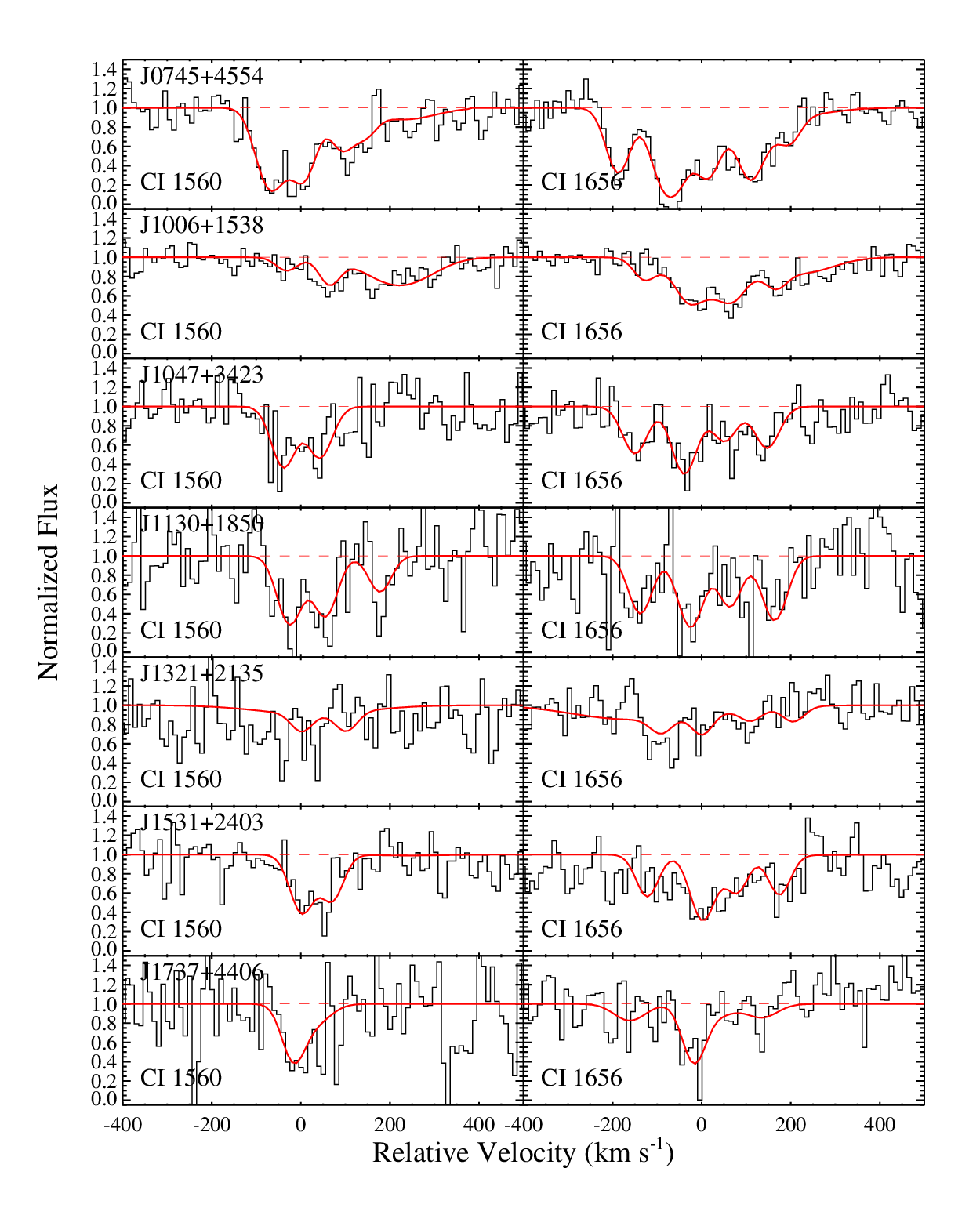}}  
\caption{C\I{} $\lambda$1560, 1656 profile fitting on the Keck/ESI spectra using VPFIT. }
\label{fig:CI}
\end{figure*}

\begin{figure}
{\includegraphics[width=8.6cm, height=6.3cm]{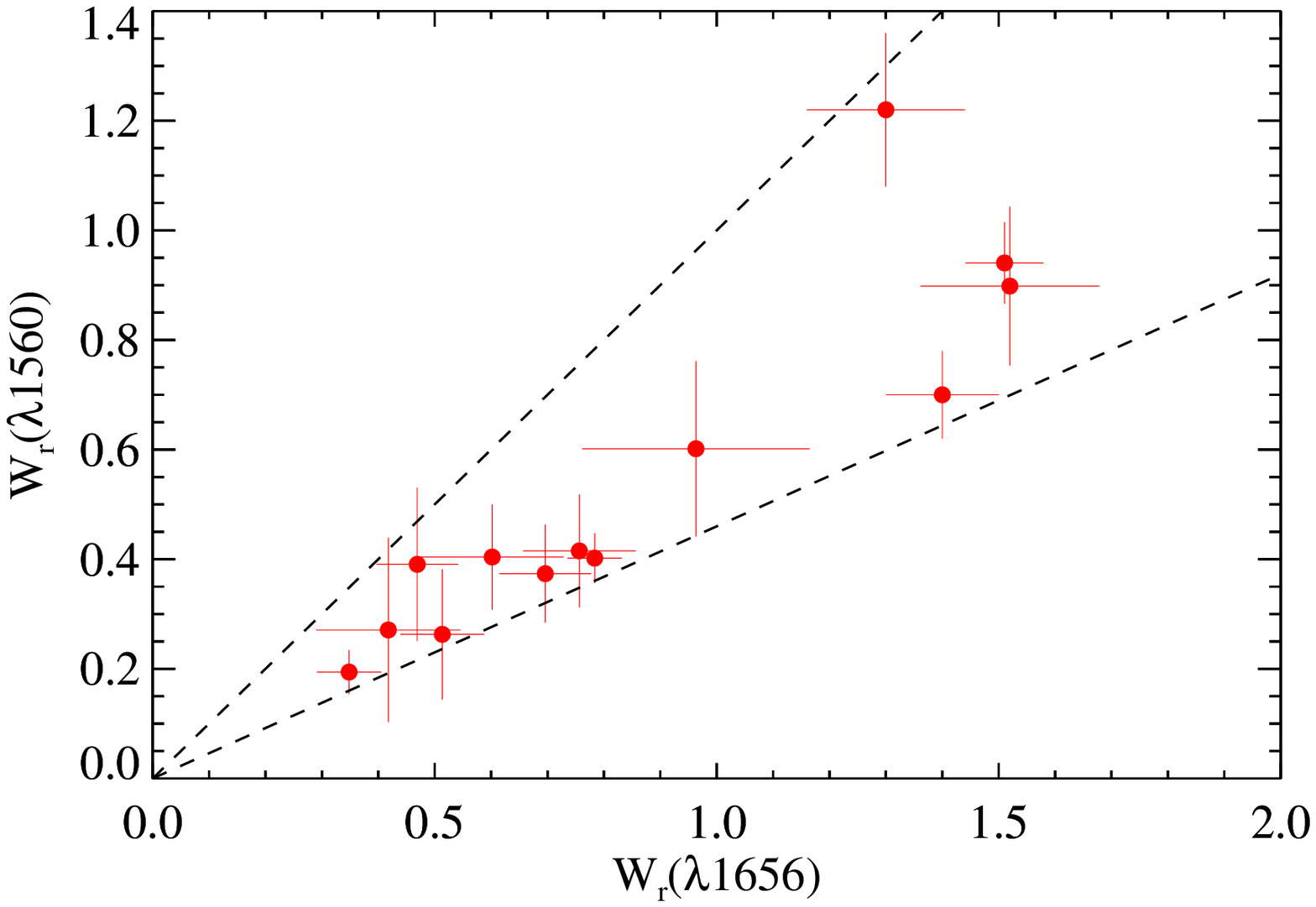}} 
\caption{W$_r$($\lambda$1560) vs. W$_r$($\lambda$1656). The upper dashed line represents the 1:1 relation expected if the two lines are heavily saturated while the lower dashed line denotes the optically thin regime.}
\label{fig:CI_EW}
\end{figure}

All the 2DAs have C\I{} detections provided their redshifts are sufficiently high to be covered in the Keck or MMT spectra. The Blue Channel Spectrograph on MMT can observe the C\I{} $\lambda$1560, 1656 lines down to $z$ $\sim$ 1.05 while Keck spectra can only probe C\I{} systems at $z$ $>$ 1.5. Figure \ref{fig:CI} shows the C\I{} velocity profiles of the Keck/ESI spectra. We measure the rest-frame equivalent widths of the C\I{} $\lambda$1560, 1656 lines by integrating the best-fit C\I{} profiles from VPFIT\footnote{http://www.ast.cam.ac.uk/~rfc/vpfit.html} assuming that the C\I{} ground state is solely responsible for all the absorption \citep{Ledoux2015}, although C\I$^*$ and C\I$^{**}$ also contribute. Figure \ref{fig:CI_EW} presents the rest-frame equivalent widths of the C\I{} $\lambda$1560, 1656 lines measured in Keck and MMT spectra. The upper dashed line represents the 1:1 relation expected if the two lines are heavily saturated, while the lower dashed line denotes the optically thin regime. All the 2DAs are located within the boundaries defined by the optically thin regime and the heavily saturated regime. We use the C\I{} equivalent widths as a measure of the amount of neutral carbon in the absorbers instead of column densities to facilitate comparison with the literature on C\I{} absorbers where only equivalent widths are reported (see Section \ref{sec:ci}).

\section{Correlation analysis}
\label{section5}

\subsection{Metallicity vs. logN(H\I)}

\begin{figure}
{\includegraphics[width=8.5cm, height=6.3cm]{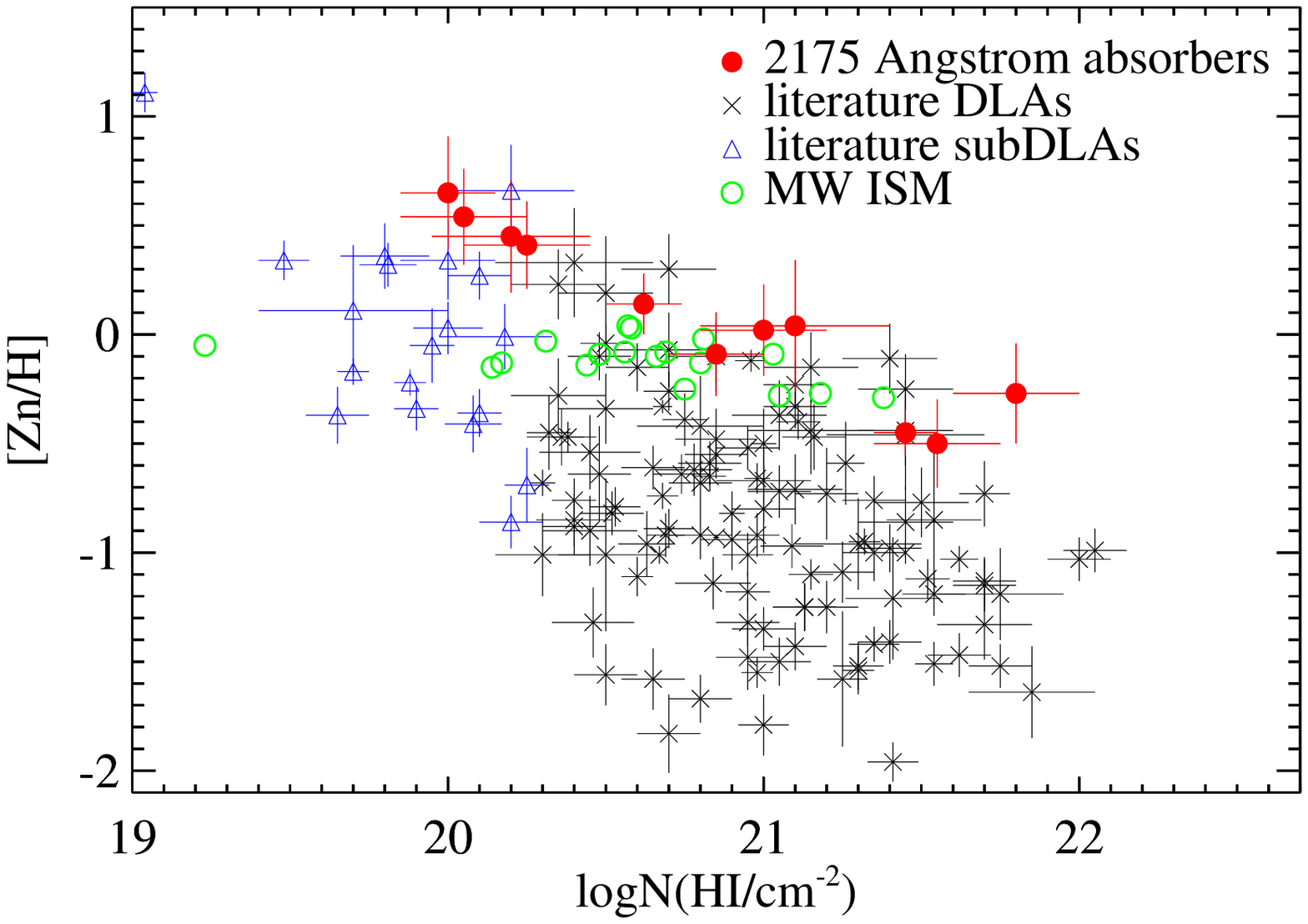}}  
\caption{[Zn/H] vs. logN(HI). The red circles are the 2175 \AA{} dust absorbers in this sample. The crosses and triangles are the literature DLAs and subDLAs with [Zn/H] measurements. The green circles are the MW clouds by \citealt{Roth1995}. }
\label{fig:ZnH_logNHI}
\end{figure}

Throughout this work, we use Zn as the metallicity indicator. Zinc is chosen as it is a non-refractory element and barely depleted. Zn has two transitions at 2026 and 2062 \AA$ $ that are rarely saturated and almost always lie redward of the Lyman-$\alpha$ forest. The 2DAs in our Keck+MMT sample all have these two transitions covered in the spectra. The [Zn/H] values are listed in Table \ref{tab:table1}.  Figure \ref{fig:ZnH_logNHI} compares our 2DAs with literature DLAs or subDLAs that have [Zn/H] measurements (see \citealt{Quiret2016} and references therein). We also display the measurements of the MW clouds by \cite{Roth1995} where different sightlines show roughly solar metallicity gas, contrary to the observed trend of increasing metallicity with decreasing N(H\I) in Figure \ref{fig:ZnH_logNHI}. SubDLAs have been known to have higher mean metallicity than DLAs (\citealt{Kulkarni2007, Khare2007, Meiring2009}), and the [Zn/H] measurements in Figure \ref{fig:ZnH_logNHI} demonstrate the same trend with the inclusion of our data: the mean [Zn/H] for subDLAs is approximately solar while it is sub-solar for DLAs. Recent studies of more Lyman limit systems, however, show that although there exists a population of high metallicity super Lyman limit systems (or sub-DLAs), the average metallicity of subDLAs is probably similar or even lower than that of DLAs \citep{Fumagalli2016,Lehner2016}. The 2DAs have logN(H\I) values that range from subDLAs to DLAs and have systematically higher [Zn/H] at each logN(H\I). It is also worth noting that the lowest logN(H\I) in this sample is $\sim$ 20.0 cm$^{-2}$; we do not see any neutral hydrogen column densities that are in the lower end of subDLAs. Therefore, the ionization corrections are less worrisome. The 2DAs in our sample are among the most metal-rich systems with relatively high neutral hydrogen content. 

\subsection{Metallicity vs. redshift}

\begin{figure}
{\includegraphics[width=8.5cm, height=6.3cm]{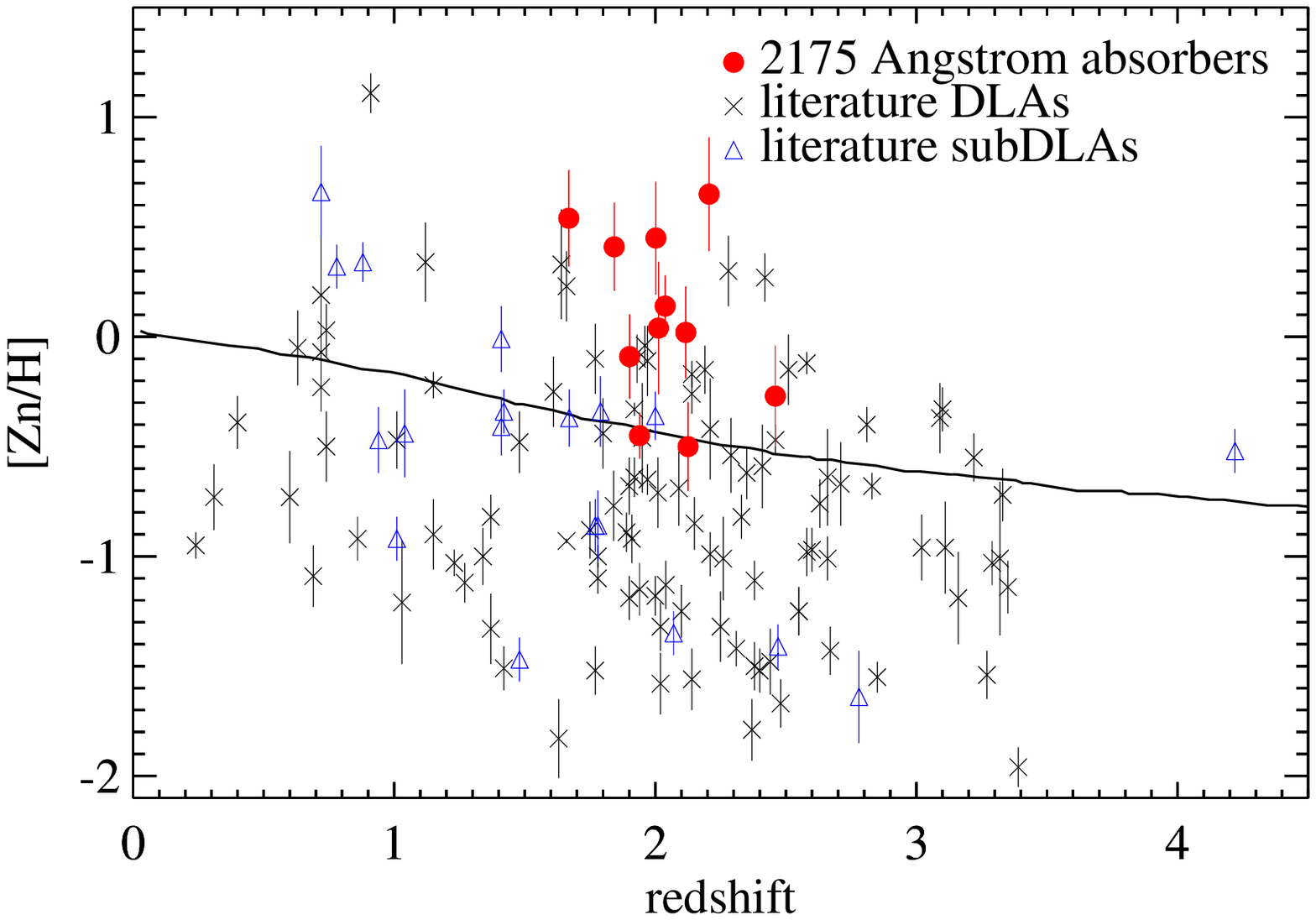}} 
\caption{[Zn/H] vs. redshift. The red circles are the 2175 \AA{} dust absorbers in this sample. The crosses and triangles are the literature DLAs and subDLAs with [Zn/H] measurements. The solid curve is the mean gas metallicity weighted by SFR from a cosmological hydrodynamic simulation investigating cosmic metal budget in various phases of baryons \citep{Dave2007}. }
\label{fig:ZnH_z}
\end{figure}

Chemical evolution models predict that global metallicity would increase over cosmic time to reach the present day solar metallicity in massive galaxies (e.g. \citealt{Lanzetta1995, Pei1995, Tissera2001}). Figure \ref{fig:ZnH_z} shows the metallicity evolution in terms of [Zn/H] as a function of redshift. The mean [Zn/H] decreases with increasing redshift as expected, while the dispersion is large at all redshifts even though the measurements are solely based on the single element, Zn. The 2DAs in our sample only cover a small range of redshifts $z$ $\sim$ 1.6 - 2.5, and we need to cover a larger redshift range to see potential evolution.  The solid curve is the mean gas metallicity weighted by SFR from a cosmological hydrodynamic simulation investigating cosmic metal budget in various phases of baryons \citep{Dave2007}. The SFR-weighted metallicity can be interpreted as a representative of $L^*$ galaxies \citep{Menard2009}. Almost all the 2DAs lie above this line, suggesting that they may be more luminous than $L^*$ galaxies. Only by detecting these galaxy counterparts in emission can we confirm or reject this speculation.

\subsection{Metallicity vs. [Fe/Zn]}

\begin{figure}
{\includegraphics[width=8.5cm, height=6.3cm]{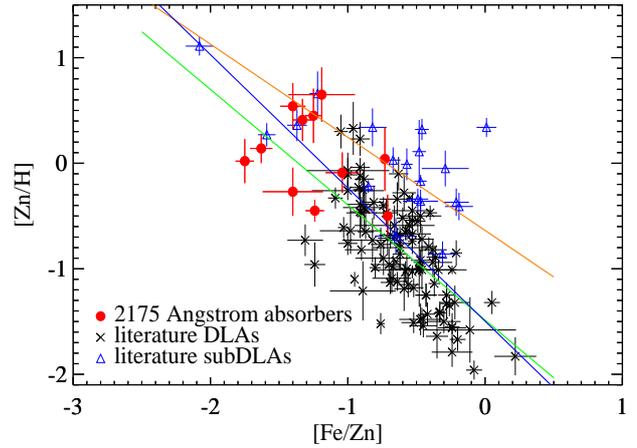}} 
\caption{[Zn/H] vs. [Fe/Zn]. The red circles are the 2175 \AA{} dust absorbers in this sample. The crosses and triangles are the literature DLAs and subDLAs with [Zn/H] and [Fe/Zn] measurements. The blue line is a linear fit to all the data points. The green line represents the linear relation for all the DLAs, and the orange line is a linear fit to all the subDLAs.}
\label{fig:ZnH_FeZn}
\end{figure}

The amount of dust in quasar absorbers is best estimated by ratios of refractory elements to volatile elements. Refractory elements more easily condense onto dust grains than volatile elements.  We measure the Fe-to-Zn gas phase abundance ratio, [Fe/Zn], to infer the dust depletion level.  In Paper I, we compare [Fe/Zn] between our 2DAs and DLA/subDLA systems in a histogram, demonstrating that the 2DAs, although a small sample, distribute at lower [Fe/Zn] (or higher depletion level) than most of DLAs and subDLAs in the literature. The high depletion level confirms the existence of dust grains and the 2175 \AA$ $ bump. We further examine [Fe/Zn] versus metallicity, which are known to be anti-correlated: high metallicity corresponds to high depletion, as shown in Figure \ref{fig:ZnH_FeZn}. The 2DAs fill the space on the [Zn/H] vs. [Fe/Zn] plane where higher [Zn/H] and lower [Fe/Zn] are expected. Dividing the sample into DLAs and subDLAs, the correlations can be fit with two lines. For DLAs, the relation (green) can be described as 
\begin{equation}
{\rm [Zn/H]} = (-1.093\pm0.053) \times {\rm [Fe/Zn]} + (-1.487\pm0.036) 
\end{equation}
\noindent with a Pearson correlation coefficient of -0.70 (10.3 $\sigma$). The subDLAs are best-fit with a line (orange) in the form of 
\begin{equation}
{\rm [Zn/H]} = (-0.883\pm0.054) \times {\rm [Fe/Zn]} + (-0.637\pm0.051) 
\end{equation}
\noindent with a Pearson coefficient of -0.72 (4.9 $\sigma$). A linear fit (blue) to all the data points yields 
\begin{equation}
{\rm [Zn/H]} = (-1.263\pm0.049) \times {\rm [Fe/Zn]} + (-1.497\pm0.035), 
\end{equation}
\noindent and the overall correlation has a larger scatter with a Pearson coefficient of -0.65 (9.8 $\sigma$).

\subsection{Metallicity vs. kinematics}

\begin{figure}
{\includegraphics[width=8.5cm, height=6.3cm]{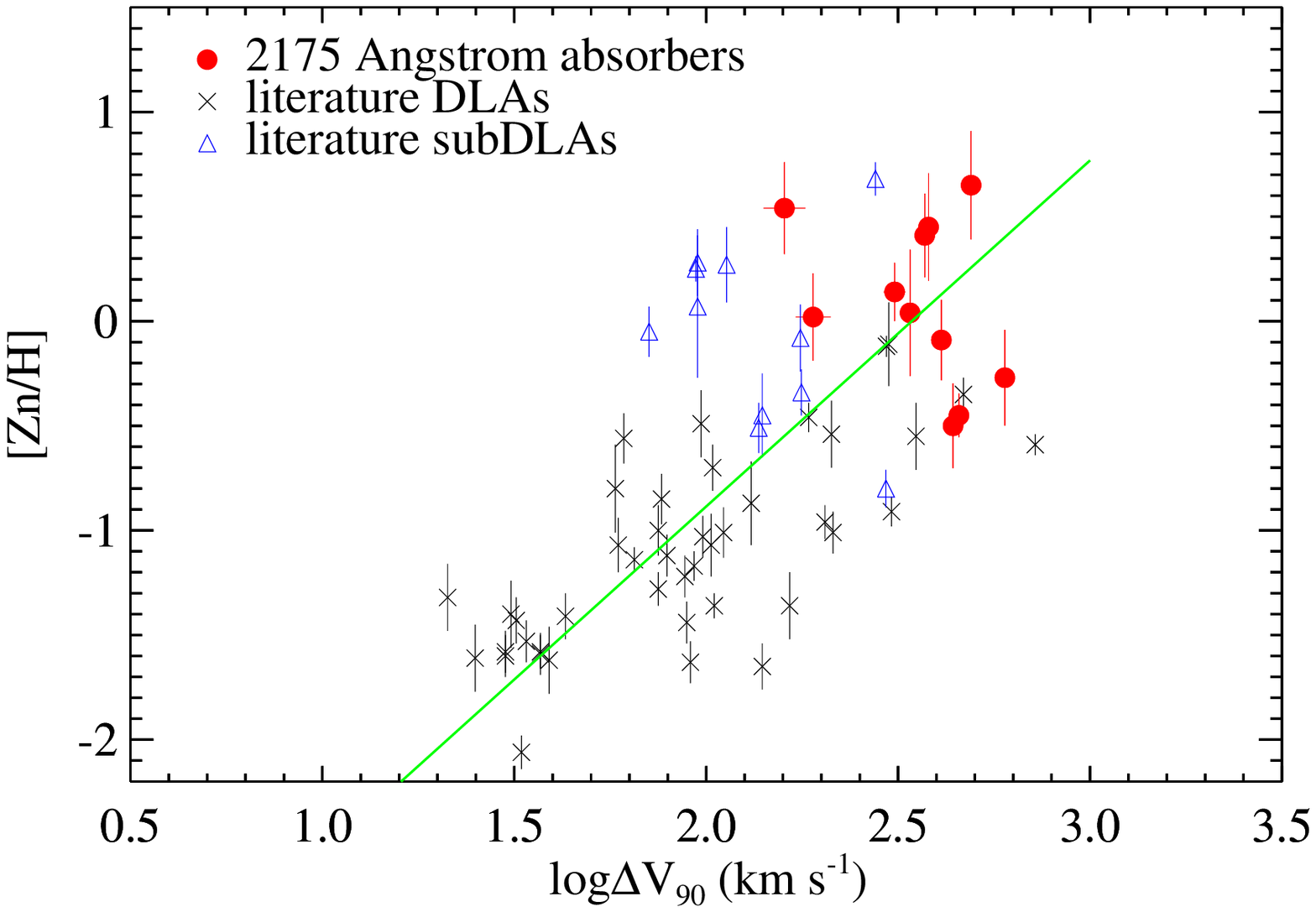}} 
\caption{[Zn/H] vs. log$\Delta$V$_{90}$. The red circles are the 2175 \AA{} dust absorbers in this sample. The crosses and triangles are the literature DLAs and subDLAs with [Zn/H] and log$\Delta$V$_{90}$ measurements. The green line is a linear fit to all the data points.}
\label{fig:ZnH_v90}
\end{figure}

A correlation between metallicity and velocity width or dispersion, $\Delta$V$_{90}$ \citep{Prochaska1997}, derived from velocity profiles has been found in DLAs as well as in subDLAs (e.g. \citealt{Ledoux2006, Meiring2007, Som2015, Quiret2016}), although subDLAs are less correlated compared to DLAs \citep{Som2015, Quiret2016}. This correlation is often interpreted as the natural product of the well-established mass-metallicity relation found at both low and high redshifts \citep{Lequeux1979, Tremonti2004, Savaglio2005, Neeleman2013} if the velocity width, $\Delta$V$_{90}$, is a reliable tracer of stellar mass.  Some simulations indeed suggest that the velocity width is driven by the gravitational potential well of the absorber's host galaxy (e.g. \citealt{Prochaska1997, Haehnelt1998, Moller2013}). 

We present [Zn/H] versus log$\Delta$V$_{90}$ in Figure \ref{fig:ZnH_v90} for literature DLAs and subDLAs that have both [Zn/H] and $\Delta$V$_{90}$ measurements. Our measurements for the velocity widths of the 2DAs are based on Fe\II{} lines, as Fe\II{} is uniformly detected in all absorbers with multiple transitions and we can choose the ones that are not blended nor saturated. The data indeed show a strong correlation with a Pearson correlation coefficient of 0.66 (7.2 $\sigma$). Our 2DAs again appear at the high end of this correlation. Overall this correlation can be fit as a straight line in the form of 
\begin{equation}
{\rm [Zn/H]} = (1.654\pm0.054) \times \log\Delta V_{90} + (-4.193\pm0.115). 
\end{equation}
\noindent Along the line of mass-metallicity-velocity width, the 2DAs are expected to have higher mean stellar mass than that of literature DLAs/subDLAs.  Together with the evidence in the previous sections, one would suspect that 2DAs are a subgroup of DLA/subDLA galaxies that have high-metallicity, high dust depletion, and are more massive/luminous.

\subsection{Mass-metallicity relation}

The correlations between metallicity and other fundamental galaxy properties such as stellar mass, luminosity, and star formation rate are crucial to understanding galaxy populations at both low and high redshifts (e.g., \citealt{Tremonti2004, Savaglio2005, Maiolino2008, Mannucci2010}), including DLA galaxies (e.g., \citealt{Ledoux2006,Fynbo2008, Pontzen2008, Prochaska2008, Krogager2012, Moller2013}).

\cite{Moller2013} analyze the redshift evolution of the mass-metallicity relation in a sample of 110 DLAs and report a formula (Equation 6) for estimating stellar mass given metallicity and redshift. \cite{Christensen2014} further test this relation by measuring the stellar masses of 12 galaxies in confirmed DLA absorber-galaxy pairs and find an excellent agreement over three orders of magnitude. They also introduce the impact parameter dependence on metallicity to the relation, which reduces the scatter in stellar masses.

However, due to the short wavelength of the Lyman-$\alpha$ absorption line, H\I{} column density measurements are usually not available for 2DAs. Since we have found strong linear correlations between [Zn/H] versus [Fe/Zn], and [Zn/H] versus log$\Delta$V$_{90}$, we can directly link stellar mass to [Fe/Zn] and log$\Delta$V$_{90}$ by propagating the correlations to the mass-metallicity relation in \cite{Moller2013} or \cite{Christensen2014}, and use [Fe/Zn] and/or log$\Delta$V$_{90}$ whichever available as rough estimators of stellar mass. Since the galaxy impact parameter is not available yet for our 2DAs, we use Equation 6 in \cite{Moller2013} assuming there is no correction from emission metallicity to absorption metallicity. The mass - log$\Delta$V$_{90}$ relation can be expressed as 
\begin{equation}
\log(M_*/M_{\odot}) = (2.911\pm0.095)$ $ \log\Delta V_{90} + 0.616 z + (1.490\pm0.202)
\end{equation}
\noindent Using the overall linear fit on the [Zn/H] - [Fe/Zn] relation, we obtain a mass - [Fe/Zn] relation in the form of 
\begin{equation}
\log(M_*/M_{\odot}) = (-2.222\pm0.086)$ $ {\rm [Fe/Zn]} + 0.616 z + (6.236\pm0.062)
\end{equation}
The stellar masses of the 2DAs based on these relations have an average (median) stellar mass of $\sim$ 2 $\times$ 10$^{10}$ M$_{\odot}$, with the highest one being an order of magnitude higher and the lowest one being a factor of $\sim$ 20 lower. The median stellar mass is comparable with that of the UV selected star-forming galaxies in a similar redshift range by \cite{Buat2012} where they find that the 2175 \AA{} bump is securely detected in 20\% of these UV selected galaxies. Unlike most DLA host galaxies, which have been very difficult to detect in emission, the host galaxies of 2DAs are expected to be detected relatively easier. These mass estimates can be used for preparing for future observations to reveal the host galaxies in emission.

\subsection{Dust-to-gas relation}

\begin{figure}
{\includegraphics[width=8.5cm, height=6.3cm]{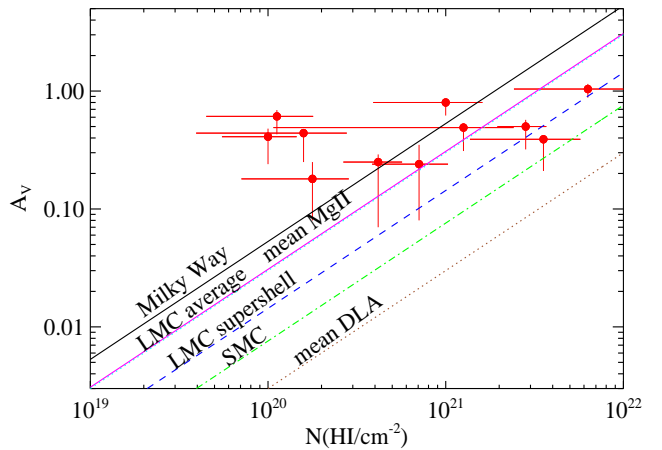}} 
\caption{Av vs. N(H\I). The red circles are the 2175 \AA$ $ absorbers in this work. The lines represent the average ratios in the MW, LMC, Mg\II{} absorbers, LMC supershell, SMC, and DLAs.  }
\label{fig:Av_NHI}
\end{figure}

The dust-to-gas ratio is one of the fundamental properties of the interstellar and intergalactic media. The ratio between visual extinction A$_{V}$ or E(B-V) to N(H\I) has been used as an estimate of dust-to-gas ratio. The presence of dust in intervening Mg\II{} and Ca\II{} absorbers has been shown in many studies \citep{Menard2003, Wang2004, York2006, Wild2006}, including a few studies reporting 2175 \AA$ $ bumps in Mg\II{} systems \citep{Wang2004,Jiang2010a,Jiang2010b,Jiang2011}. The 2DAs in our sample are also pre-selected from Mg\II{} absorber catalogs. \cite{Menard2009} find the mean dust-to-gas ratio of the Mg\II{} absorbers is $\langle A_{V}\rangle$/$\langle N(\rm HI) \rangle$ = (3.0 $\pm$ 0.6) $\times$ 10$^{-22}$ mag cm$^2$.  The A$_{V}$/N(H\I) of MW is a factor of $\sim$ 1.8 higher \citep{Bohlin1978}. The average A$_{V}$/N(H\I) of the LMC is comparable with that of Mg\II{} absorbers, while the LMC supershell is a factor of $\sim$ 2 lower than the average value \citep{Gordon2003}. The dust-to-gas ratio in the SMC is almost an order of magnitude lower than that in the MW \citep{Gordon2003}. \cite{Vladilo2008} report a mean $\langle A_{V}\rangle$/$\langle N(\rm HI) \rangle$ for 250 DLAs at  2.2 $<$ $z$ $<$ 3.5 of $\sim$ 2 - 4 $\times$ 10$^{-23}$ mag cm$^2$, an order of magnitude lower than the mean dust-to-gas ratio of Mg\II{} absorbers. Figure \ref{fig:Av_NHI} presents the A$_{V}$-N(H\I) relation for the 2DAs, which cross the space from LMC supershell to MW. A few of them are higher in A$_{V}$ than that of MW at the same N(H\I). The average (median) dust-to-gas ratio for this sample is $\sim$ 6.0 $\times$10$^{-22}$ mag cm$^2$, which is consistent with that of the local ISM (4-6 $\times$10$^{-22}$ mag cm$^2$; \citealt{Liszt2014}). The absence of a strong correlation here is likely because we are selecting based on dust, i.e., there is a lower limit to $A_V$ imposed which prevents us from including many systems with low N(H\I) and low $A_V$.

\begin{figure}
{\includegraphics[width=8.5cm, height=6.3cm]{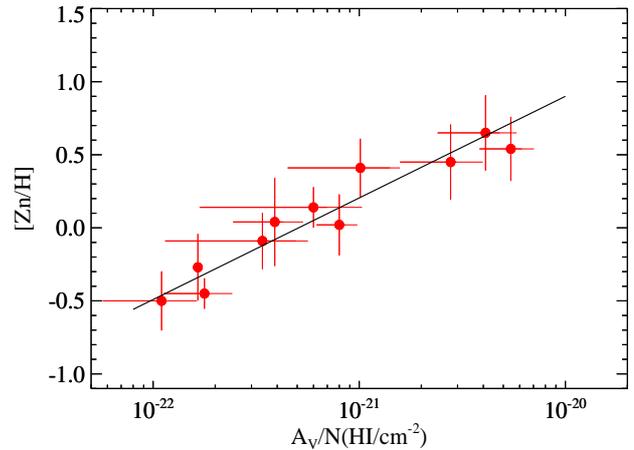}} 
\caption{[Zn/H]  vs. Av/N(H\I). The dust-to-gas ratio is proportional to metallicity. }
\label{fig:ZnH_AvNHI}
\end{figure}

In statistical studies (e.g. \citealt{Menard2009} where individual metallicity measurements are missing, their analysis is based on the assumption that the dust-to-gas ratio on average is proportional to metallicity. Since we have [Zn/H] measurements for every 2DA, we present the data on the plane of metallicity versus dust-to-gas ratio in Figure \ref{fig:ZnH_AvNHI}, which exhibits a strong linear correlation. The linear relation can be expressed as 
\begin{equation}
{\rm [Zn/H]} = 0.695 $ $ \log(A_{\rm V}/N({\rm H~{\sc I}})) + 14.8.  
\end{equation}
\noindent Our data confirm the scenario that dust forms out of metals; actually this relation is a proxy for metal-to-dust ratio yet another way of presenting it. A universal metal-to-dust ratio constant to within a factor of 30-40\% has been found in quasar absorbers and GRB afterglows, which implies the dominant formation mechanism of the bulk of the cosmic dust  \citep{Zafar2013}.  However, the formation mechanism of the 2175 \AA$ $ bump carriers is still an open question.

\section{Relationship with other absorption line systems}
\label{section6}

2DAs are a new population of quasar absorption line systems whose role in the context of absorbers and galaxy populations has not been established. Based on the properties that we have investigated, we explore how 2DAs are related to other absorption systems. The relationship also reveals the conditions that are required to produce the 2175 \AA$ $ bump.

\subsection{2DAs are Mg\II{} and Fe\II{} absorbers}

All the 2DAs selected from the SDSS surveys are pre-identified as Mg\II{} and Fe\II{} absorbers. Therefore they are all Mg\II/Fe\II{} absorbers by the selection criterion. This procedure could cause a selection bias towards Mg\II/Fe\II{} absorbers. However, 2DAs without Mg\II{} and Fe\II{} lines are not expected since Mg and Fe are among the abundant elements responsible for the material of the 2175 \AA$ $ feature \citep{Draine2003}. Whether non-Mg\II/Fe\II{} absorbers also contain the 2175 \AA$ $ absorption feature can be tested by developing searching algorithms that do not depend upon the pre-selection criterion, i.e., applying deep learning techniques \citep{Yuan2016}. 

\subsection{2DAs are metal-strong DLAs/subDLAs}

We have examined all the systems whose redshifts are sufficiently high for Lyman-$\alpha$ absorption line to be covered in the spectra, and the H\I{} column density measurements indicate that they are either DLAs or subDLAs, i.e., high H\I{} gas content. All of them exhibit logN(H\I) $\geq$ 20.0 thus far. In Paper I, we also compare the [Fe/Zn] and log$\Delta$V$_{90}$ histograms between the 2DAs and general DLAs/subDLAs as well as metal-strong DLAs (MS-DLAs; logN(Zn\II) $\geq$ 13.15 or logN(Si\II) $\geq$ 15.95) defined by \cite{Herbert-Fort2006}. The KS tests suggest that the 2DAs are more likely to be drawn from the same parent population as metal-strong DLAs/subDLAs than normal DLAs/subDLAs. All but one 2DAs from Paper I and this work satisfy the quantitative criterion on N(Zn\II) or N(Si\II). \cite{Kaplan2010} examine the stacked quasar spectrum of $\sim$ 40 MS-DLAs selected from the SDSS DR5. A possible broad absorption line feature is visible at $\sim$ 2135 \AA$ $ but their fits (SMC-like vs. MW-like extinction curve) are inconclusive. It would be interesting to search for bumps in all the metal-strong DLAs/subDLAs from the literature to test if the inverse statement is also true.

\subsection{2DAs are C\I{} absorbers}
\label{sec:ci}

\begin{figure}
{\includegraphics[width=8.5cm, height=6.3cm]{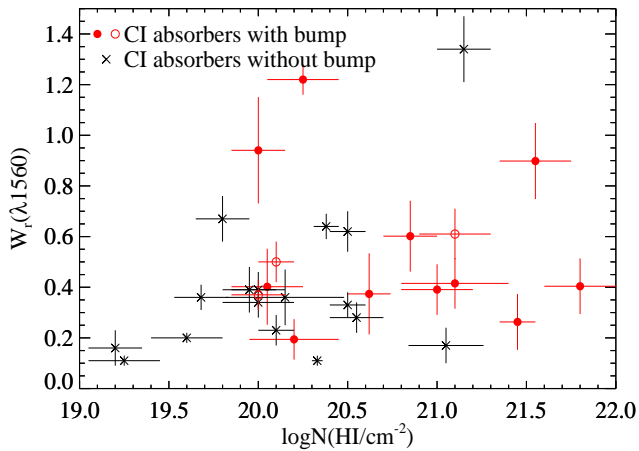}}  
\caption{W$_r$($\lambda$1560) vs. logN(H\I). The red filled circles are the 2175 \AA{} dust absorbers in this sample and the unfilled circles are the C\I{} absorbers with a 2175 \AA{} extinction feature (\citealt{Ledoux2015}). The crosses are C\I{} absorbers without a 2175 \AA{} bump. }
\label{fig:wr1560_logNHI}
\end{figure}

We adopt the rest-frame equivalent width of C\I{} $\lambda$1560 as an estimate of the amount of neutral carbon in the absorbers. In Figure \ref{fig:wr1560_logNHI}, we plot the equivalent width of C\I{} $\lambda$1560 versus logN(H\I) for our 2DAs (with both C\I{} and H\I{} coverage) and the C\I{} absorbers with N(H\I) measurements from \cite{Ledoux2015}. They find that the fraction of C\I{} absorbers among subDLAs is much less than among DLAs (see Figure 9 in \citealt{Ledoux2015}), suggesting that high N(H\I) is preferred to provide efficient shielding for C\I{} to exist. For the C\I{} absorbers with relatively low hydrogen column densities (i.e. 19.0 $<$ logN(H\I) $<$ 20.3), dust is expected to be present to provide extra shielding by absorbing UV photons. As shown in Figure \ref{fig:wr1560_logNHI}, the C\I{} absorbers with the 2175 \AA$ $ extinction feature (i.e. our 2DAs plus three C\I{} absorbers in \citealt{Ledoux2015}) have higher N(H\I) and higher W$_r$($\lambda$1560) on average than the ones without the 2175 \AA$ $ bump, although the correlation between W$_r$($\lambda$1560) and logN(H\I) is weak and the samples with and without the 2175 \AA$ $ bump are both small and incomplete. 

About 30\% of the C\I{} absorbers in \cite{Ledoux2015} are 2DAs. We further plot W$_r$($\lambda$1560) versus bump strength for the 2DAs from the C\I{} sample together with the 2DAs in this work. Although the data do not indicate a strong linear correlation between W$_r$($\lambda$1560) and bump strength, the equivalent widths for all the 2DAs are above $\sim$ 0.2 \AA$ $ and the C\I{} column densities are all above $\sim$10$^{14}$ cm$^{-2}$. The 2DAs discovered in GRB afterglows also contain C\I{}, and the measured equivalent widths of C\I{} and bump strengths suggest larger equivalent widths for stronger bumps albeit of a small sample \citep{Zafar2012}. All of the 2DAs being C\I{} absorbers may be generalized that 2DAs are a subset of C\I{} absorbers, implying that the physical and chemical conditions that favor C\I{} are required to yield the 2175 \AA$ $ bump.

\begin{figure}
{\includegraphics[width=8.5cm, height=6.3cm]{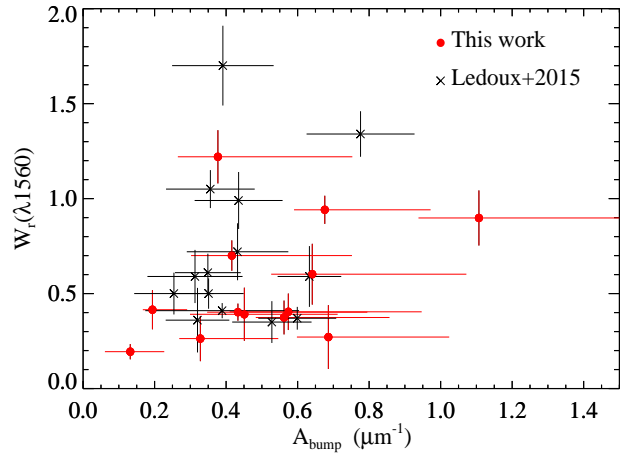}}  
\caption{W$_r$($\lambda$1560) vs. $A_{\rm bump}$. The red filled circles are the 2175 \AA{} dust absorbers in this sample and the crosses are the C\I{} absorbers from \citealt{Ledoux2015} that contain a 2175 \AA{} bump. }
\label{fig:wr1560_abump}
\end{figure}

\subsection{2DAs as molecular gas tracers}

The simultaneous presence of the 2175 \AA$ $ bump and molecular gas ($H_2$ and CO) has been reported in a few absorbers \citep{Noterdaeme2009,Prochaska2009,Ma2015,Noterdaeme2017}. In searching for $H_2$ in DLAs, high metallicity is an important criterion given the correlation between metallicity and the depletion of metals onto dust grains, i.e., larger amount of $H_2$ is expected in higher metallicity DLAs which contain more dust \citep{Ge1997,Ge2001,Petitjean2006,Noterdaeme2008}. The low detection rate is partly due to the fact that these molecular absorption lines are often buried in the Lyman-$\alpha$ forest, inhibiting a detection with a high signal-to-noise ratio. More importantly, the cold molecular gas can be missed due to relatively small cross-sections compared to the more pervasive warm neutral ISM \citep{Zwaan2006}. Neutral carbon has been considered an excellent tracer of molecular hydrogen ($H_2$) because the ionization potential of C\I{} (11.26 eV) is very close to the dissociation energy of $H_2$ (e.g., \citealt{Srianand2005}). When $H_2$ is detected, C\I{} is always present in the same system (e.g.,\citealt{Ge1999,Ge2001,Srianand2008,Guimaraes2012,Noterdaeme2015,Balashev2017}.) Neutral Chlorine (Cl\I), although rarely detected, is an alternative excellent tracer of $H_2$ \citep{Jura1974,Moomey2012,Noterdaeme2010,Ma2015,Balashev2015}. 

The production of CO becomes significant when C\I{} dominates the carbon species and a large portion of hydrogen turns molecular \citep{Snow2006}. Targeting strong C\I{} absorbers leads to more successful detections \citep{Noterdaeme2009,Noterdaeme2010,Noterdaeme2011,Balashev2017} after the first detection at high-redshift by \cite{Srianand2008}. However, this condition is not sufficient to obtain CO in detectable amounts because CO requires more shielding from UV photons than $H_2$ and C\I{} given a dissociation energy of 11.09 eV. The additional shielding provided by dust grains is critical, therefore targeting dusty systems like our 2DAs should lead to a higher detection rate of CO. The required conditions (high metallicity, high depletion, etc.) for the bump carriers to exist also facilitate the production of molecular gas. 2DAs are expected to be efficient molecular gas tracers.

\subsection{Intervening and proximate systems}
\label{sec:proximate}

Most of the 2DAs are intervening systems along the sight lines of the background quasars but physically unrelated to quasars. There are three systems with velocities relative to the quasar emission redshifts that are less than \mbox{5000 \kms} therefore referred to as ``proximate systems", which could be associated with the quasar or the quasar's host galaxy. The residual Ly$\alpha$ emission in the absorption trough towards J1047+3423 can be interpreted as associated with star formation activity or scattered Ly$\alpha$ photons from the quasar (e.g., \citealt{Noterdaeme2014,Pan2017}). For two of them (J1006+1538, J1705+3543), their absorption redshifts are even higher than their emission redshifts by up to $\sim$ 3000 \kms. These systems could be explained by infalling gas to the quasar or other mechanisms \citep{Lu2007}. The presence of the 2175 \AA$ $ bump near the quasar environment may have implications on the production and survival of the dust grains responsible for the bump. However, we do not observe particularly different properties of the 2175 \AA$ $ proximate absorbers with respect to the intervening systems. These systems are of interest on their own and we may investigate them in greater detail that is beyond the scope of this paper.

\section{Conclusions}
\label{section7}

We have studied the neutral content (H\I{} and C\I) of 13 2DAs at $z$ = 1.6 - 2.5 selected from SDSS/BOSS and followed up with Keck and MMT. We derive the absolute metal abundances and perform a correlation analysis between metallicity and H\I{} column density, redshift, depletion level, and velocity width. Based on the well-established mass-metallicity relation, we also derive a formula for estimating stellar masses by using alternative tracers, i.e., [Fe/Zn] or $\Delta$V$_{90}$. We have explored the relationship between 2DAs and other absorption line systems (DLAs/subDLAs, Mg\II/Fe\II{} absorbers, C\I{} absorbers). 

\begin{itemize}

\item The 2DAs in this sample are among the highest metallicity compared to literature DLAs/subDLAs and MW ISM, and they all have relatively high H\I{} content. The 2DAs occupy the region with both high metallicity and high depletion levels on the [Zn/H] vs. [Fe/Zn] plane. They lie at the high end of the [Zn/H]-log$\Delta$V$_{90}$ correlation.

\item The correlation between [Zn/H] and  [Fe/Zn] or [Zn/H] and log$\Delta$V$_{90}$ can be used as alternative stellar mass estimators based on the well-established mass-metallicity relation. The estimated stellar masses of the 2DAs in this sample are in the range of $\sim$10$^9$ to 2 $\times$ 10$^{11}$ $M_{\odot}$ with a median value of $\sim$2 $\times$ 10$^{10}$ $M_{\odot}$. 

\item The relationship with other quasar absorption line systems can be described as (1) 2DAs are a subset of Mg\II{} and Fe\II{} absorbers, (2) 2DAs are preferentially metal-strong DLAs/subDLAs, (3) More importantly, all the 2DAs show C\I{} detections, (4) 2DAs can be used as molecular gas tracers. Three 2DAs in this work are proximate systems although we do not observe distinct properties compared to the intervening absorbers.

\end{itemize}

Combined with the analyses in \cite{Ma2015} and Paper I, we have confirmed a correlation between the presence of the 2175 \AA$ $ bump and other properties including high metallicity, high depletion, large velocity width, overall low-ionization state of the gas, and simultaneous presence of neutral and molecular gas. The 2DAs appear to trace metal-rich galaxies with plenty of dust, neutral, and molecular gas for star formation at high redshifts. Their host galaxies are likely to be chemically enriched, evolved, massive (more massive than DLA/subDLA galaxies), and presumably star-forming galaxies although we have not made direct measurements for the stellar mass and star formation rate yet. We expect them to be detected in emission more easily than DLA hosts. It is very promising that atomic and molecular emission lines as well as dust continuum can be detected with ALMA in the host galaxies of 2DAs.

\section*{Acknowledgments}

We thank Tayyaba Zafar for the constructive comments which have improved the paper. This work was partially supported by the University of Florida and the UF-UCF SRI program. Jingzhe Ma was funded by the UF alumni fellowship. This work has made use of data obtained by the SDSS-I/II, SDSS-III/BOSS, Keck, and MMT.  

Funding for the SDSS and SDSS-II has been provided by the Alfred P. Sloan Foundation, the Participating Institutions, the National Science Foundation, the U.S. Department of Energy, the National Aeronautics and Space Administration, the Japanese Monbukagakusho, the Max Planck Society, and the Higher Education Funding Council for England. The SDSS Web Site is http://www.sdss.org/. The SDSS is managed by the Astrophysical Research Consortium for the Participating Institutions. The Participating Institutions are the American Museum of Natural History, Astrophysical Institute Potsdam, University of Basel, University of Cambridge, Case Western Reserve University, University of Chicago, Drexel University, Fermilab, the Institute for Advanced Study, the Japan Participation Group, Johns Hopkins University, the Joint Institute for Nuclear Astrophysics, the Kavli Institute for Particle Astrophysics and Cosmology, the Korean Scientist Group, the Chinese Academy of Sciences (LAMOST), Los Alamos National Laboratory, the Max-Planck-Institute for Astronomy (MPIA), the Max-Planck-Institute for Astrophysics (MPA), New Mexico State University, Ohio State University, University of Pittsburgh, University of Portsmouth, Princeton University, the United States Naval Observatory, and the University of Washington.

Funding for SDSS-III has been provided by the Alfred P. Sloan Foundation, the Participating Institutions, the National Science Foundation, and the U.S. Department of Energy Office of Science. The SDSS-III web site is http://www.sdss3.org/. SDSS-III is managed by the Astrophysical Research Consortium for the Participating Institutions of the SDSS-III Collaboration including the University of Arizona, the Brazilian Participation Group, Brookhaven National Laboratory, Carnegie Mellon University, University of Florida, the French Participation Group, the German Participation Group, Harvard University, the Instituto de Astrofisica de Canarias, the Michigan State/Notre Dame/JINA Participation Group, Johns Hopkins University, Lawrence Berkeley National Laboratory, Max Planck Institute for Astrophysics, Max Planck Institute for Extraterrestrial Physics, New Mexico State University, New York University, Ohio State University, Pennsylvania State University, University of Portsmouth, Princeton University, the Spanish Participation Group, University of Tokyo, University of Utah, Vanderbilt University, University of Virginia, University of Washington, and Yale University.

The W.M. Keck Observatory is operated as a scientific partnership among the California Institute of Technology, the University of California and the National Aeronautics and Space Administration. The Observatory was made possible by the generous financial support of the W.M. Keck Foundation.  

The 6.5m MMT telescope is operated by the MMT observatory, a joint venture of the smithsonian institution and the University of Arizona.

\bsp	
\label{lastpage}
\end{document}